# Ecosystem Recovery to Historical Targets Becomes Unattainable Under Modelled Fishing and Climate in the Barents Sea

Matthew Hatton[1], Jack H Laverick[1], Neil Banas[1], Michael Heath[1]

[1]Department of Mathematics and Statistics, University of Strathclyde, Glasgow, United Kingdom

Alternative titles:

Historical Recovery Targets Become Unattainable Under Climate Change in the Barents Sea

Declining Ecosystem Resilience in the Barents Sea Under Climate and Harvest Stress

Ecosystem Recovery After Harvest Stress is Challenged by Climate Change in the Barents Sea.

Correspondence:

Matthew Hatton,

Department of Mathematics and Statistics, University of Strathclyde, Glasgow, United Kingdom

Email: matthew.hatton@strath.ac.uk

## Abstract

Climate change and fisheries jointly shape the resilience of the Barents Sea marine ecosystem, yet the recovery of key fish populations to climate and anthropogenic disturbances requires further investigation. This study examines how fishing pressure and climate change, driven by the NEMO-MEDUSA Earth system model, influence the recovery times of Demersal and Planktivorous fish in the Barents Sea. We used the StrathE2EPolar end-to-end ecosystem model to simulate transient dynamics under increasing fishing pressure scenarios, and quantified recovery times for Demersal, Planktivorous, and ecosystem-wide groups relative to a shifting unfished baseline. Recovery times increased with both fishing intensity and climate change, by as much as 18 years for Demersal fish and 54 years for Planktivorous fish across all fishing scenarios. At the ecosystem level, recovery was constrained by the slow rebound of top predators, many of which experienced biomass collapse under climate change, preventing recovery to a shifting baseline. Our results suggest that fishing pressure in tandem with climate change substantially reduces ecosystem resilience, highlighting the importance of sustainable harvest strategies in a changing climate.

# 1 Introduction

Demersal and Planktivorous fish species represent key ecological and economic components of the Barents Sea ecosystem (Stokke et al. 1999; Bogstad et al. 2015; ICES 2021). Target Demersal species such as Haddock, Saithe, and Polar Cod support major commercial fisheries (Olsen et al. 2010), with 2023 catches totalling 29 505, 2 642, and 108 070 metric tonnes respectively (Marine Stewardship Council 2023). Planktivorous fish, particularly Capelin and Herring, also underpin large-scale fisheries and are a critical prey resource for a variety of high trophic level species (Dolgov 2016; Kaartvedt and Titelman 2018).

Demersal fish, as mid trophic level consumers, form an important link between low trophic level organisms (e.g. zooplankton and Benthic Invertebrates) and high trophic level species (e.g. Polar Bears, Seabirds, and Seals) (Whitehouse et al. 2017). Similarly, Planktivorous fish act as a bridge which transfers energy from lower trophic level zooplankton to both Demersal predators and Marine Mammals (Gjøsæter 1998; Rose 2005). As a result, fluctuations in both Demersal and Planktivorous fish biomass not only affects top predators but can also cause wider ecosystem impacts down the food web (Ripple et al. 2016), making each species valuable indicators for ecosystem-based management (Frank et al. 2005; Johannesen et al. 2012).

The region's relatively well-monitored Demersal and Planktivorous stocks, which are jointly managed by Norway and Russia (Stokke et al. 1999), focus on precautionary management strategies to aid sustainable exploitation of key species (Gjøsæter et al. 2012). However, even in this highly regulated system, both species are subject to a range of pressures that extend beyond direct harvesting (Emblemsvåg et al. 2022).

Despite global efforts (Arctic Council 2015), the Barents Sea is experiencing a changing climate (Årthun et al. 2025), with potentially negative consequences for Demersal and Planktivorous stocks (Richardson 2008; Reygondeau and Beaugrand 2011; Eriksen et al. 2025). Rising seawater temperatures, reduced sea ice cover, and migrating fish communities are altering food web dynamics and species distributions (Hansen et al. 2024). Arctic Cod and Capelin have expanded their range northwards (Huse and Ellingsen 2008; Roderfeld et al. 2008; Fall et al. 2018; Ingvaldsen et al. 2021), and some arctic ecosystems have been introduced to new pathogens, parasites, and non-indigenous species (Huserbråten et al. 2019; Chan et al. 2019; Årthun et al. 2025). These shifts may disrupt established predator-prey relationships, increase competition from boreal generalists, and ultimately undermine the stability and productivity of key fish populations (Kortsch et al. 2015).

Consequently, changes in Demersal and Planktivorous fish abundance or ecosystem structure, whether due to fishing pressure, climate change, or their coupled effects, all propagate through the ecosystem (Ripple et al. 2016). Understanding the resilience of Demersal and Planktivorous stocks, their capacity to recover following environmental or anthropogenic perturbations (Holling 1973; Walker et al. 2006), is critical for informing adaptive management and conservation strategies under future climate conditions (Cochrane et al. 2009; Thorpe et al. 2023).

Stock recovery is a key goal under sustainable management frameworks (Lynam et al. 2016). The Marine Stewardship Council (MSC) defines ‘less sensitive’ habitats as those that would be able to recover to at least 80% of their unimpacted structure and function within 20 years, assuming fishing ceases entirely (Council 2022). This recovery benchmark is used as a threshold for determining whether a fishery causes ‘serious or irreversible harm’, potentially deeming a fishery uncertifiable. However, under intensifying climate change and shifting ecosystem baselines (Pauly 1995; Hobday 2015; Alleway et al. 2023), it is unclear whether such a threshold will remain ecologically realistic or operationally useful in the future.

One study by Thorpe, Heath, and Lynam (2023) explored methods for estimating recovery times of different functional guilds and examined how climate variability may influence these trajectories in the North Sea. Building on this foundation, the present study incorporates both fishing pressure scenarios and long-term climate change to simulate future conditions in the Barents Sea. In doing so, we aim to assess the applicability of the widely used MSC certification benchmark in a changing Arctic environment (Jørgensen et al. 2022; Smith et al. 2025).

Here, we assess how fishing scenarios and climate affect the recovery times of Demersal and Planktivorous fish stocks, as well as ecosystem-wide metrics using the end-to-end ecosystem model StrathE2EPolar (Heath et al. 2022). StrathE2E models have been widely applied to explore food web interactions and ecosystem responses to environmental and anthropogenic pressures (Michael Heath, Robert Wilson and Douglas Speirs 2015; Heath et al. 2022; Thorpe et al. 2022; Laverick et al. 2025), but have yet to be used to assess the transient dynamics of key fish stocks alongside ecosystem-wide recovery. This study aims to fill the gap in understanding how climate changes and fishing pressures may alter recovery times in the Barents Sea. We seek to answer the following research questions:

- How do the combined effects of climate change and fishing pressures affect the recovery trajectories, timescales, and resilience of Demersal and Planktivorous fish stocks in the Barents Sea?

- How do ecosystem-wide recovery times compare to those of target fish stocks, and what does this reveal about the effectiveness of single-guild recovery as a proxy for ecosystem health?
- How do recovery criteria behave under future climate and harvest conditions?

# 2 Materials and Methods

All of the following data manipulation and modelling were conducted in the R programming language (R version 4.3.1, RDC Team (2023)) using RStudio (Team 2020). The study used several R packages to analyse the data. Specifically, `dplyr' (Wickham 2015) for data manipulation and cleaning and `sf' (Pebesma 2018) to encode spatial data. All plots were visualised using the `ggplot2' package (Wickham 2016). All code is available on GitHub.

## 2.1 Model Description

Box 1, StrathE2E Marine Ecosystem Model Description:

StrathE2E combines a dynamic model of marine ecology and a static model of fishing activity, coded as a package for the R computing environment (Heath et al. 2021). The ecology model is a set of coupled ordinary differential equations representing the continuous-time rates of change in nitrogen mass of 26 non-living and living guilds of material due to flows through a food web network. The guilds span dissolved inorganic nutrient, detritus and microbes, through classes of plankton, benthos and fish to birds and mammals. The flows represent predation, food assimilation, metabolism, excretion, demographic transitions (for fish and benthos guilds; spawning and recruitment), passive advection, active migrations, and fishery harvest. The spatial resolution is coarse, in keeping with the guild granularity of the food web - an "inshore" and "offshore" zone, linked by advection and migration. Each zone is further subdivided into 4 seabed habitats and, in the offshore zone, 2 water column layers. The seabed habitats are biogeochemical compartments with discrete rates of processing detritus mineralisation and nutrient recycling constrained by sediment grain size, and the sensitivity of these to disturbance e.g. by fishing abrasion. External environmental drivers are monthly resolution annual cycles of time-varying physical and chemical data (temperatures, marine hydrodynamics and river hydrology, sea surface light, turbidity, and boundary data on inorganic nutrient concentrations and volumetric fluxes of water). The model outputs state variable nitrogen masses of each guild in each compartment and the flows between them, at daily intervals.

The StrathE2E fishing model integrates externally defined properties of up to 12 fishing fleets - groups of vessels defined by the fishing gears they use - which selectively harvest distinctive sets of guilds in the ecology model. Properties of each fleet are their habitat-level spatial distributions of activity rate, selectivity for ecology model guilds, discarding, offal generation and seabed abrasion rates. The outputs from the fishing model are integrated zone and guild-level fishing mortality, discarding and offal generation rates, and habitat-level seabed abrasion rates. These are injected as parameters into the ecology model and assumed to remain constant over all upcoming years of simulation.

The StrathE2EPolar model builds on the existing temperate shelf-sea fisheries food web model outlined in Box 1. Arctic and Antarctic ecosystems present unique challenges, in part due to their dynamic sea-ice coverage (Windnagel et al. 2017; Chavas and Grainger 2019), extreme seasonal variations (Vader et al. 2025), and the ecological importance of ice-dependent species (Callaghan et al. 2004; Marz 2010; Kayode-Edwards et al. 2024). As polar regions are the first to feel the effects of climate change (Ritchie et al. 2021; Kayode-Edwards et al. 2024; Hatton et al. 2025), it has become essential to incorporate these dynamics into existing ecosystem models.

StrathE2EPolar extends StrathE2E by incorporating additional environmental drivers, such as time varying ice concentration and thickness as well as snow dynamics, allowing for a representation of polar marine food webs. Such dynamics will affect the habitable area by arctic specific guilds such as Maritime mammals, while also attenuating the light reaching the water column (Laverick et al. 2025).

StrathE2EPolar is available as an R package (https://marineresourcemodelling.gitlab.io/sran/index.html) and contains an example implementation for the Barents Sea. Details on the parametrisation process of the Barents Sea model are found within Heath et al. (2022).

***Table 1 Guilds in StrathE2EPolar.*** *The left column presents a list of frequently discussed guilds within this study. The right column gives examples of which species may be contained within each guild. This list is not exhaustive but aims to guide the reader on which species are considered to be contained within StrathE2EPolar's guilds.*

| **Guild** | **Example of Species within the Model Domain** |
|---|---|
| Demersal Fish | Atlantic Cod, Haddock, Halibut, Saithe |
| Planktivorous Fish | Blue Whiting, Capelin, Forage fish, Polar Cod |
| Maritime Mammals | Polar Bears |
| Cetaceans | Orca, Northern Bottlenose Whale, Humpback Whales |
| Pinnipeds | Harp Seal, Hooded Seal, Walrus |
| Birds | Auks, Cormorants, Gannets, Petrels, Terns |

## 2.2 Model Inputs

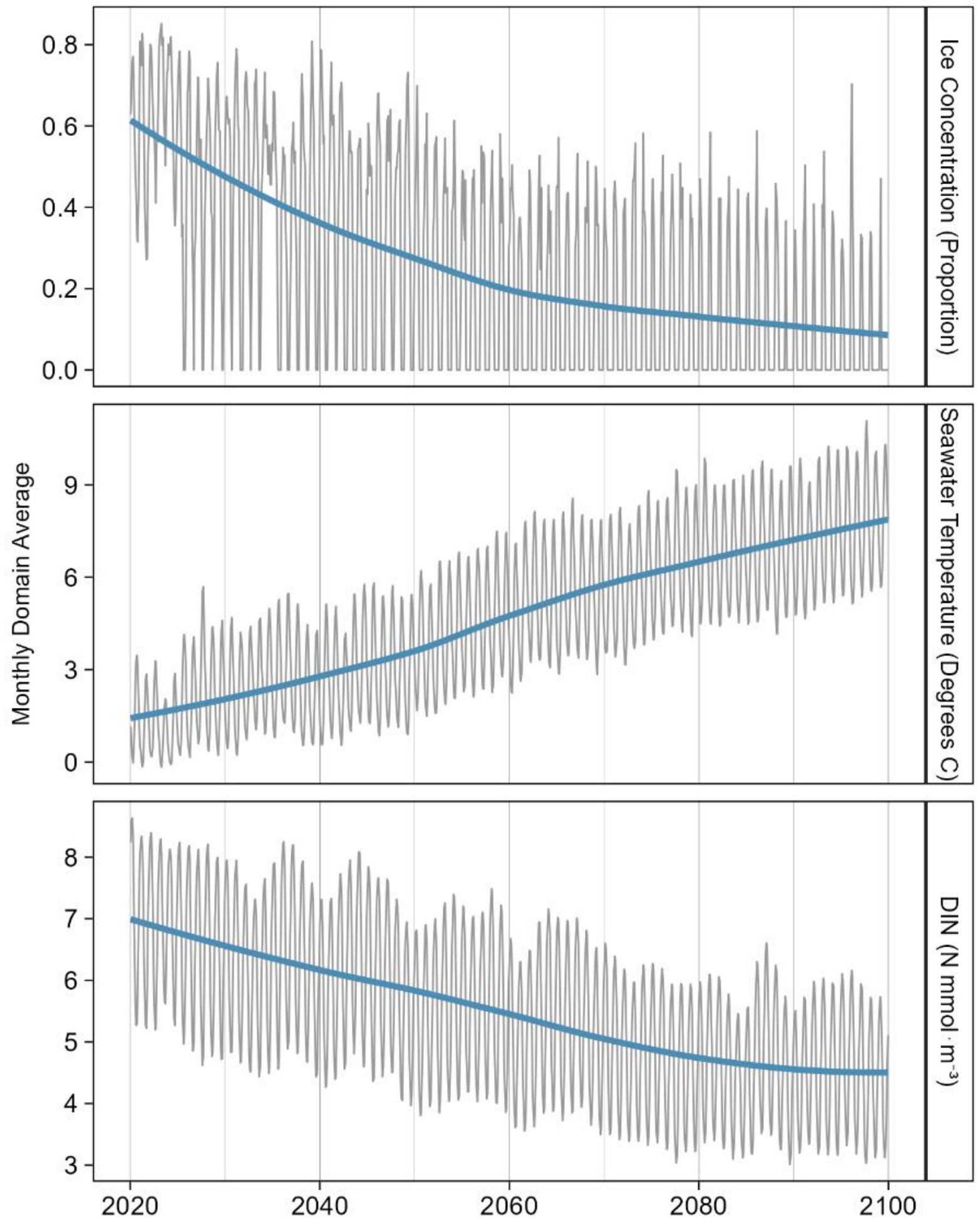

***Figure 1 Climate Drivers**. Three key climate drivers used to force transient dynamics for the Barents Sea StrathE2EPolar model. Decreasing trends in Ice Concentration (Proportion) and DIN (Dissolved Inorganic Nutrients; millimoles of Nitrogen per meter cubed) are shown, whilst an increasing trend in Seawater Temperature (Degrees Celsius) persists throughout the century.*

The StrathE2EPolar model was driven by time-varying physical and biogeochemical inputs derived from the NEMO-MEDUSA Earth system model (NM; Yool, Popova, and Anderson 2013). NM was configured at a quarter-degree resolution and run from 1980 to 2099 under the RCP8.5 greenhouse gas emissions scenario (Riahi et al. 2011).

Environmental drivers, including seawater temperature, ice/snow concentration/thickness, light, riverine inputs, vertical mixing, and nutrient concentrations, were extracted from NM for the Barents Sea model domain (see Figure 1 in Heath et al. (2022)). Monthly time series of driving data for each ocean volume were produced using a 10-year smoothing window, yielding a sequence of annually varying environmental drivers from 2020 to 2099. Climate change in the Barents Sea entails increasing ocean temperatures and decreases in nutrient and sea-ice concentrations (Figure 1). A full list of environmental inputs and their processing methods can be found in Heath et al. (2022).

## 2.3 Transient Simulations

To simulate transient ecosystem dynamics, the model was initialised with conditions from year $t$, forced with year $t + 1$ driving data, and run for one year. The resulting state variables were extracted and used to initialise the next year's simulation, allowing for stepwise progression through the changing climate. We refer to this scheme as StrathE2EPolar's transient mode (Figure 2).

## 2.4 Fishing Scenarios and Recovery

To define a meaningful recovery state, we computed the unfished ecosystem trajectory as a transient baseline. We ran the model in transient mode without any fishing pressure, allowing ecosystem state variables to evolve under progressive environmental forcing (Figure 2). This captured momentum and inertia in the ecosystem as climate conditions change. This baseline was used to calculate a dynamic MSC 80% threshold per year, which provided a condition for evaluating ecosystem recovery.

We conducted separate transient simulations including fishing of Demersal and Planktivorous fish. Fishing was simulated at nine time points evenly spaced through-out the century from 2020 to 2085. The fished model for each year was run to a steady state using a 50-year burn-in. After the burn-in period all fishing pressure was removed (Figure 2) and the system ran in transient mode until the end of the century. Recovery times were calculated for each simulation as the number of years required for the biomass of the targeted guild to return to 80% of the unfished baseline value, consistent with the MSC threshold. The recovery time for each guild in StrathE2EPolar was also calculated per simulation. The slowest recovery time across all guilds was used as a measure of ecosystem-wide recovery.

To investigate the effect of fishing intensity on recovery times we defined three scenarios. We first estimated the maximum sustainable yield (MSY) for Demersal and Planktivorous fish by generating yield curves (Supplementary Figures 1 and 2). MSY was identified as the maximum landings from each curve, taking then the harvest rate that achieves MSY. Based on these results we defined three fishing scenarios applying harvest ratio multipliers that scale the relative fishing effort exerted on each guild. The three scenarios were:

1. **Status Quo** - a baseline multiplier of 1x, corresponding to a present-day low-effort regime which is approximately 44% of MSY for Demersal fish (Hvingel and Zimmermann 2023) and 78% for Planktivorous fish (Heath et al. 2022).
2. **$MSY_{2020}$** - effort scaled to achieve a maximum sustainable yield for the target guild in 2020. Harvest ratio multipliers of 2.8 for Demersal fish and 1.5 for Planktivorous fish.

3. **Overfishing** - a high-intensity scenario applying twice the $MSY_{2020}$ effort, representing overexploitation. Harvest ratio multipliers of 5.6 for Demersal fish and 3 for Planktivorous fish.

To isolate the effects of pressure on each guild, harvest ratio multipliers were applied only to the focal group (Demersal or Planktivorous fish), with all other harvest ratios set to zero during the simulation.

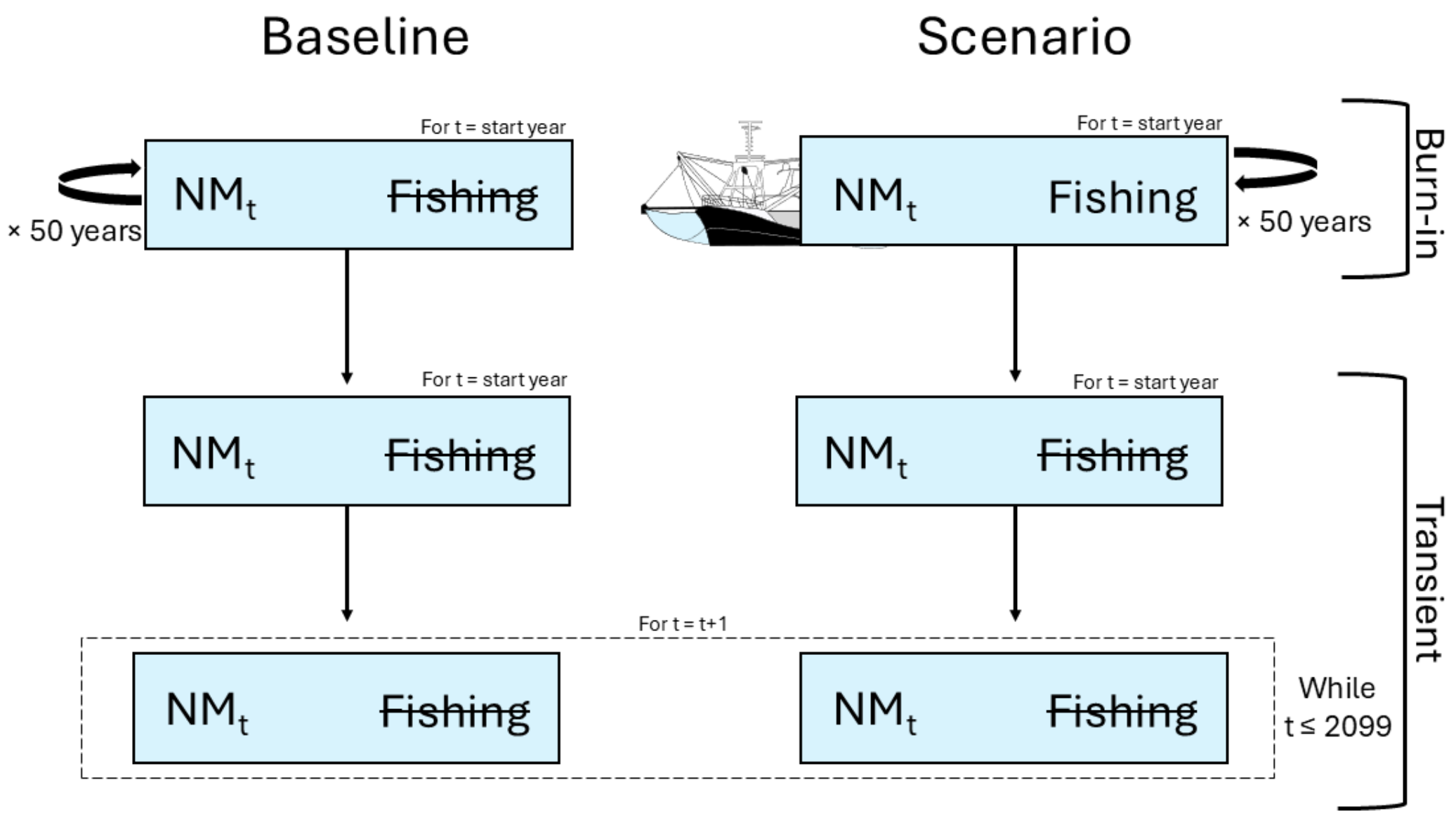

***Figure 2 The Experimental Design.*** *A flowchart showing the experimental design for this study. For each fished year between 2020 and 2085 fishing pressure was applied independently for Demersal and Planktivorous fish as a Scenario. An experiment was launched with a 50-year burn-in model run. After the burn-in period, the transient run was launched with all fishing pressure removed. During the transient period the environmental forcings were updated annually. The baseline scenario was calculated in the same way but using an unfished burn-in period. NM represents the NEMO-MEDUSA driving data for year t.*

## 2.5 Casual Inference

Using the same method as Laverick et al. (2025), we conducted a set of experiments to explore the causes of change in the unfished baseline. In each experiment one group of driving data (Table 2) were held at their initial values while all other drivers followed projected climate trajectories.

*Table 2: Definitions of experimental groups. This table is a reproduction of the one found in* Laverick et al. (2025)*. Note: SO, SI, and D indicate the surface offshore, surface inshore, and deep model compartments respectively. For each listed experiment in the table, the drivers mentioned were held at the values for the 2010-2020 smoothed period; all other drivers were updated year-on-year.*

| Experiment | Drivers Held Constant |
| --- | --- |
| Nitrogen Concentrations | Concentrations of nitrogen mass sources at the model boundary<br>• SO_nitrate<br>• SO_ammonia<br>• SO_phyt<br>• SO_detritus<br>• D_nitrate<br>• D_ammonia<br>• D_phyt<br>• D_detritus<br>• SI_nitrate<br>• SI_ammonia<br>• SI_phyt<br>• SI_detritus |
| Flows | Water volume exchanges for all model compartments<br>• SO_OceanIN<br>• D_OceanIN<br>• SI_OceanIN<br>• SI_OceanOUT<br>• SO_SI_flow |
| Ice | Variables related to the StrathE2EPolar cryosphere module<br>• SO_IceFree<br>• SI_IceFree<br>• SO_IceCover<br>• SI_IceCover<br>• SO_IceThickness<br>• SI_IceThickness<br>• SO_SnowThickness<br>• SI_SnowThickness |
| Light | Surface Irradiance<br>• SLight |
| Temperature | Air and Ocean Temperature for all model compartments<br>• SO_temp |
| | • D_temp<br>• SI_temp<br>• SO_AirTemp<br>• SI_AirTemp |

# 3 Results

## 3.1 Demersal Fish Biomass Trajectories

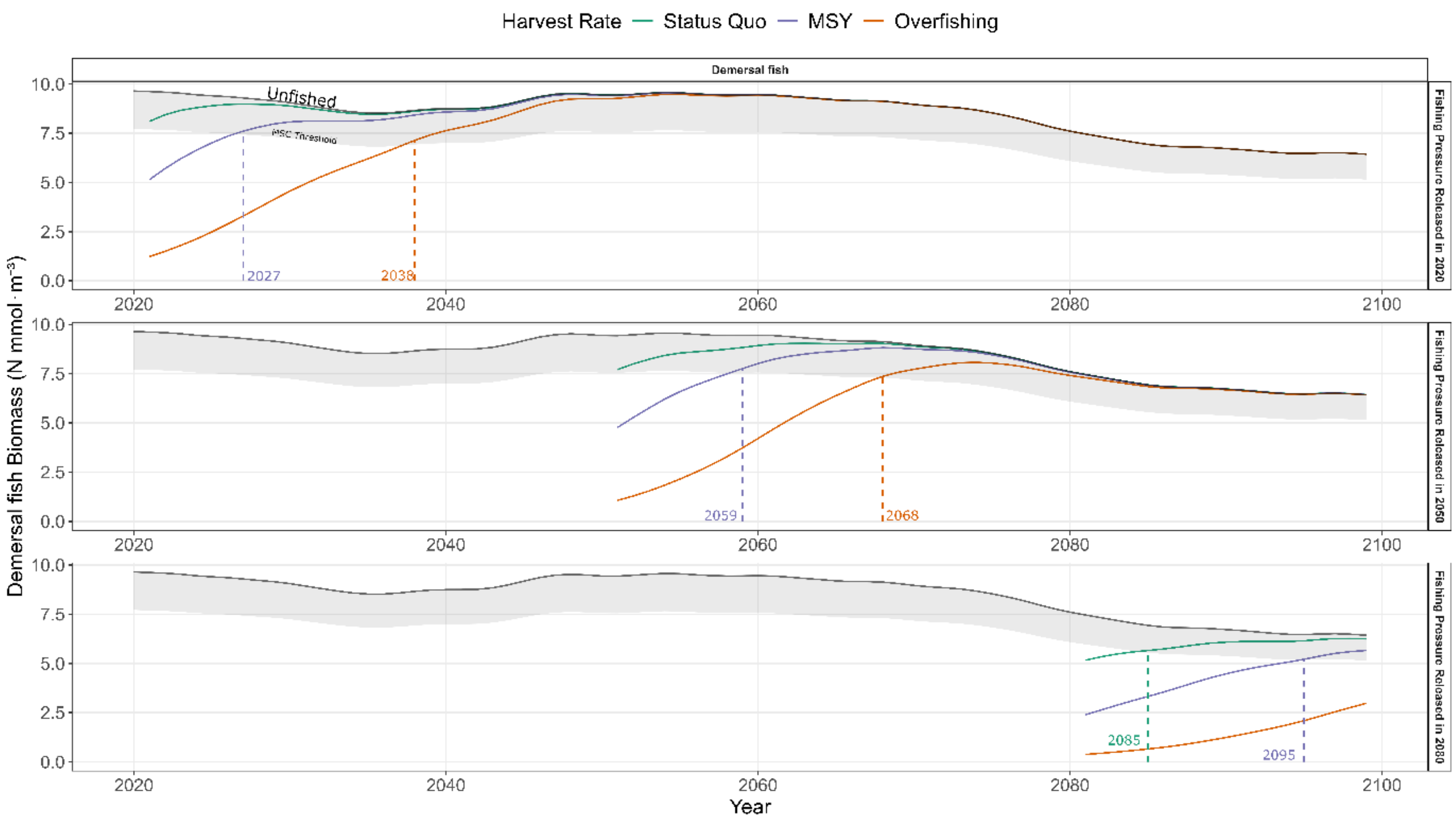

***Figure 3 Demersal Fish Biomass Trajectories.*** *Demersal fish biomass trajectories for 3 distinct fishing release years, 2020, 2050, and 2080. The solid black line indicates the unfished baseline Demersal fish biomass throughout the century. The dark-grey ribbon below the solid black line indicates 80% of the unfished biomass, in line with the MSC threshold. Recovery trajectories are shown for each of the three fishing scenarios, Status Quo (green),* $MSY_{2020}$ *(blue), and Overfishing (orange). The corresponding recovery year is marked with a dashed line in the same colour for each scenario. The units of fish biomass are millimoles of Nitrogen per meter cubed within the model domain. Recovery trajectories for the full set of release years can be found in Supplementary Figure 3.*

For Demersal fish biomass, the unfished transient baseline remained relatively steady between 9.65 and 8.75 ($N\ mmol \cdot m^{-3}$) during early-mid-century decades (2020–2060) but exhibited a gradual decline beyond 2060 in response to changing climate (Figure 3). This decline became more apparent under late-century (2070-2085) climate forcings, with unfished biomass decreasing by up to 32.6% of the 2020s value, to 6.4 $N\ mmol \cdot m^{-3}$.

In the 2020s release, status quo fishing effort caused Demersal fish biomass to decline by 16% from 9.6 to 8.1 $N\ mmol \cdot m^{-3}$. Despite the fishing pressure, the biomass remained above the 80% MSC threshold, so no recovery was required. The $MSY_{2020}$ scenario produced a larger decrease of 47.2% to 5.1 $N\ mmol \cdot m^{-3}$, with biomass recovering in 2027 following the cessation of fishing. When overfishing, the biomass dropped by 87.6% from the unfished baseline to 1.2 $N\ mmol \cdot m^{-3}$, with recovery to the MSC threshold delayed until 2038.

In the 2050s release, declines were more pronounced across all fishing scenarios. Biomass under status quo fell by 18.4% from 9.4 to 7.7 $N\ mmol \cdot m^{-3}$, remaining above the MSC threshold. The $MSY_{2020}$ scenario resulted in a decline of 49.5% to 4.8 $N\ mmol \cdot m^{-3}$, with recovery to the MSC threshold observed by 2059. Under overfishing, biomass declined to 1.1 $N\ mmol \cdot m^{-3}$ (88.3%), with recovery occurring in 2068.

In the 2080s release, biomass losses intensified, and recovery times lengthened. Under status quo, biomass fell by 31.6% from 7.6 to 5.2 $N\ mmol \cdot m^{-3}$, falling outside the MSC threshold, recovering in 2085. The $MSY_{2020}$ scenario led to a reduction of 68.4% to 2.4 $N\ mmol \cdot m^{-3}$, with recovery by 2095. Under overfishing, biomass declined to 0.4 $N\ mmol \cdot m^{-3}$ (94.7%), with recovery impossible before the end of the simulation period in 2099. Table 3 shows the percentage and absolute change in biomass, as well as the recovery times for Demersal fish.

*Table 3: Percentage, absolute change and recovery time in Demersal fish biomass under each fishing scenario, per simulation year. NR represents when the recovery time exceeded that of the remaining simulation time.*

| Scenario/Release Year | Status Quo | | | $MSY_{2020}$ | | | Overfishing | | |
|---|---|---|---|---|---|---|---|---|---|
| | Percentage Change (%) | Absolute Change ($N\ mmol \cdot m^{-3}$) | Recovery Time (years) | Percentage Change (%) | Absolute Change ($N\ mmol \cdot m^{-3}$) | Recovery Time (years) | Percentage Change (%) | Absolute Change ($N\ mmol \cdot m^{-3}$) | Recovery Time (years) |
| 2020 | 16 | 1.5 | 0 | 47.2 | 4.5 | 7 | 87.6 | 8.4 | 18 |
| 2050 | 18.4 | 1.7 | 0 | 49.5 | 4.6 | 9 | 88.3 | 8.5 | 18 |
| 2080 | 31.6 | 2.4 | 5 | 68.4 | 5.2 | 15 | 94.7 | 7.2 | NR |

## 3.2 Planktivorous Fish Biomass Trajectory

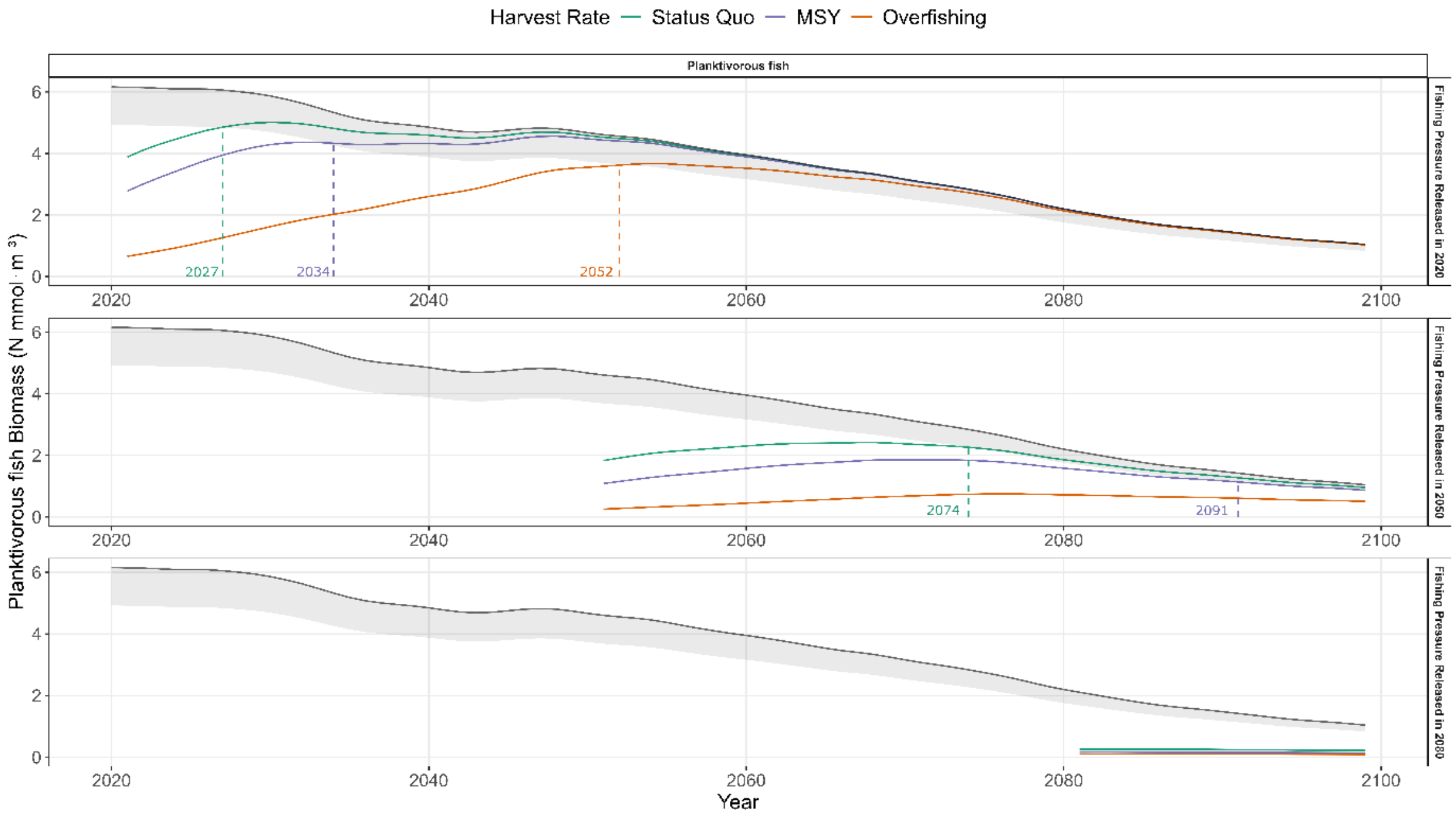

***Figure 4 Planktivorous Fish Biomass Trajectories.*** *Planktivorous fish biomass trajectories for 3 distinct fishing release years, 2020, 2050, and 2080. The solid black line indicates the unfished baseline Planktivorous fish biomass throughout the century. The dark-grey ribbon below the solid black line indicates 80% of the unfished biomass, in line with the MSC threshold. Recovery trajectories are shown for each of the three fishing scenarios, Status Quo (green), $MSY_{2020}$ (blue), and Overfishing (orange). The corresponding recovery year is marked with a dashed line in the same colour for each scenario. The units of fish biomass are millimoles of per meter cubed within the model domain. Recovery trajectories for the full set of release years can be found in Supplementary Figure 4.*

For Planktivorous fish biomass, the unfished transient baseline declined steadily throughout the century by 83.9% from 6.2 to 1 $N\ mmol \cdot m^{-3}$ due to climate change (Figure 4).

In the 2020s, Planktivorous fish biomass fell from 6.2 to 3.9 $N\ mmol \cdot m^{-3}$ under the status quo fishing scenario, a decline of 36.7%, recovering to this state in 2027. Under $MSY_{2020}$, the biomass reduction was more severe, reaching 2.8 $N\ mmol \cdot m^{-3}$ (54.5% decline), but levels rebounded to the MSC threshold by 2034 following the removal of fishing pressure. Overfishing caused a pronounced collapse to 0.7 $N\ mmol \cdot m^{-3}$, amounting to an 88.6% loss from the unfished baseline; recovery in this scenario was delayed until 2052.

By the 2050s, the impacts of fishing on the system became more apparent. Under status quo fishing pressure planktivorous biomass dropped further from 4.7 to 1.8 $N\ mmol \cdot m^{-3}$. The biomass recovered to the MSC threshold by 2074. The $MSY_{2020}$ scenario saw biomass reduce by 76.4%, reaching 1.1 $N\ mmol \cdot m^{-3}$ before recovering in 2091. Overfishing drove Planktivorous fish biomass to minimal levels from 4.7 to 0.3 $N\ mmol \cdot m^{-3}$, representing a 93.6% decrease, with recovery impossible during the remaining simulation time.

Planktivorous fish biomass declines in the 2080s were markedly more severe. Even under status quo fishing, biomass was reduced by 86.4% from 2.2 to 0.3 $N\ mmol \cdot m^{-3}$ and remained below the MSC threshold, with no recovery observed by 2099 despite the cessation of fishing pressure. The $MSY_{2020}$ scenario led to slightly further losses, reducing biomass by 90.9% to 0.2 $N\ mmol \cdot m^{-3}$; again, exhibiting no signs of recovery. Under the overfishing scenario, biomass dropped by 95.5% to 0.1 $N\ mmol \cdot m^{-3}$.

*Table 4: Percentage, absolute change, and recovery time in Planktivorous fish biomass under each fishing scenario, per simulation year. NR represents when the recovery time exceeded that of the remaining simulation time.*

| Scenario/Release Year | Status Quo | | | $MSY_{2020}$ | | | Overfishing | | |
|---|---|---|---|---|---|---|---|---|---|
| | Percentage Change (%) | Absolute Change ($N\ mmol \cdot m^{-3}$) | Recovery Time (years) | Percentage Change (%) | Absolute Change ($N\ mmol \cdot m^{-3}$) | Recovery Time (years) | Percentage Change (%) | Absolute Change ($N\ mmol \cdot m^{-3}$) | Recovery Time (years) |
| 2020 | 36.7 | 3.9 | 7 | 54.5 | 2.8 | 14 | 88.6 | 0.7 | 32 |
| 2050 | 61.7 | 1.8 | 24 | 76.4 | 1.1 | 41 | 93.6 | 0.3 | NR |
| 2080 | 86.4 | 0.3 | NR | 90.9 | 0.2 | NR | 95.5 | 0.1 | NR |

Table 4 shows the percentage and absolute change in biomass, as well as the recovery times for Planktivorous fish. Full recovery trajectories for both Demersal and Planktivorous fish across all release years (2020 to 2085) are presented in Supplementary Figures 3 and 4.

## 3.3 Demersal Fish Recovery Times

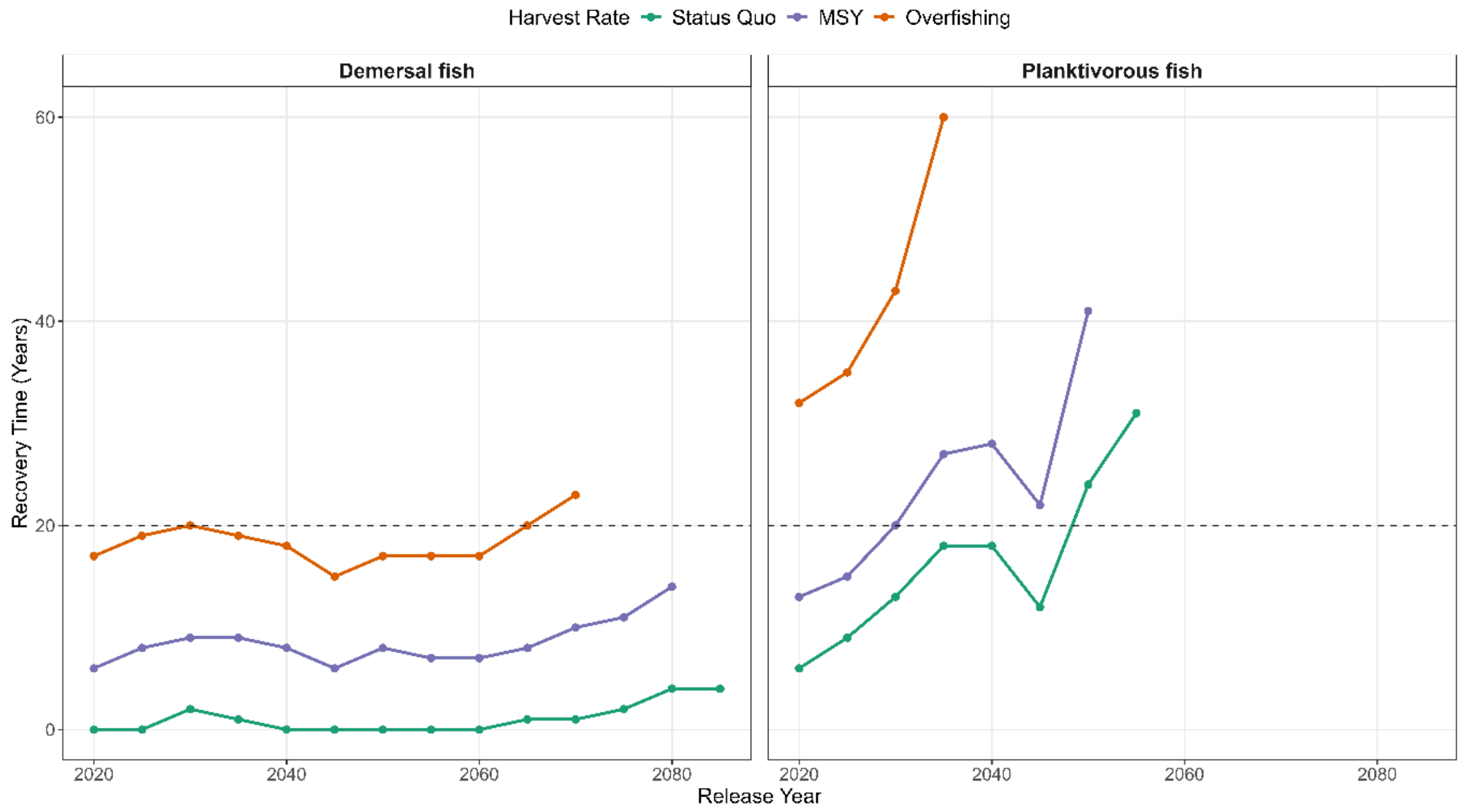

***Figure 5 Demersal and Planktivorous Fish Biomass Recovery Times**. Demersal and Planktivorous fish recovery times after fishing stops in different years between 2020 and 2085. Each fishing scenario begins from a fished steady state (50-year burn-in) and is released of all fishing pressure for a transient run. Recovery time is calculated as the difference between the release year and the first year at which the biomass crossed 80% of the unfished baseline value for each fishing scenario. The units of fish biomass are millimoles of Nitrogen per meter cubed within the model domain. The 20-year recovery (dashed line) is marked as the maximum recovery time for a fishery to be certified by the MSC.*

Recovery times for Demersal fish biomass are projected to increase under scenarios with higher fishing mortality and climate effects (Figure 5). Under status quo fishing pressure, recovery times remain negligible across most of the simulation period. Recovery times are consistently near 0 between 2020 and 2060. A brief increase is observed in 2030, with a recovery time of 2 years, and after 2060, recovery durations slightly increase to a maximum of 4 years by 2080 and 2085.

In contrast, the $MSY_{2020}$ scenario shows consistently higher recovery times while remaining below the MSC 20-year threshold. Fishing between 2020 and 2060 results in recovery times ranging between 6 and 9 years, peaking in 2030. We then see a gradual increase until 2080, where recovery times peak at 14 years. By releasing fishing pressure in 2085, the recovery time exceeds that of the remaining simulation time (14 years).

The overfishing scenario exhibits the longest recovery durations throughout each release year, with a general increasing trend over time. Recovery times begin at 17 years in 2020 and gradually lengthen, exceeding the MSC threshold of 20 years by 2070. After this point, recovery is no longer possible within the remaining simulation window.

## 3.4 Planktivorous Fish Recovery Times

Recovery times for Planktivorous fish biomass in the Barents Sea are projected to notably rise under the combined effects of climate change and fishing, frequently exceeding the MSC threshold of 20 years (Figure 5). In the status quo scenario, recovery time starts at 6 years in 2020, then increases sharply to 18 years by 2035–2040. Following this, a steep increase is seen, rising to a peak of 31 years in 2055. Beyond this point, recovery does not occur within the remaining 39-year simulation period.

Under the $MSY_{2020}$ scenario, recovery times follow a broadly similar trajectory to the status quo, though with consistently longer recovery times throughout the simulation intervals. A recovery time of 13 years is observed in 2020, rising steadily to 20 years by 2030, already breaching the MSC threshold. Following this, we see an increase beyond the MSC threshold to 28 years in 2040. In 2050, a peak of 41 years is seen, more than twice that of the MSC threshold. As with the status quo scenario, recovery then becomes unachievable within the remaining simulation period of 44 years.

In the overfishing scenario, recovery times are already well beyond the MSC threshold at the outset, starting at 32 years in 2020. A steep increase follows, with recovery time peaking at 60 years by 2035. From that point onward, recovery is no longer possible within the remaining 59-year simulation period.

## 3.4 What is driving a decrease in unfished biomass?

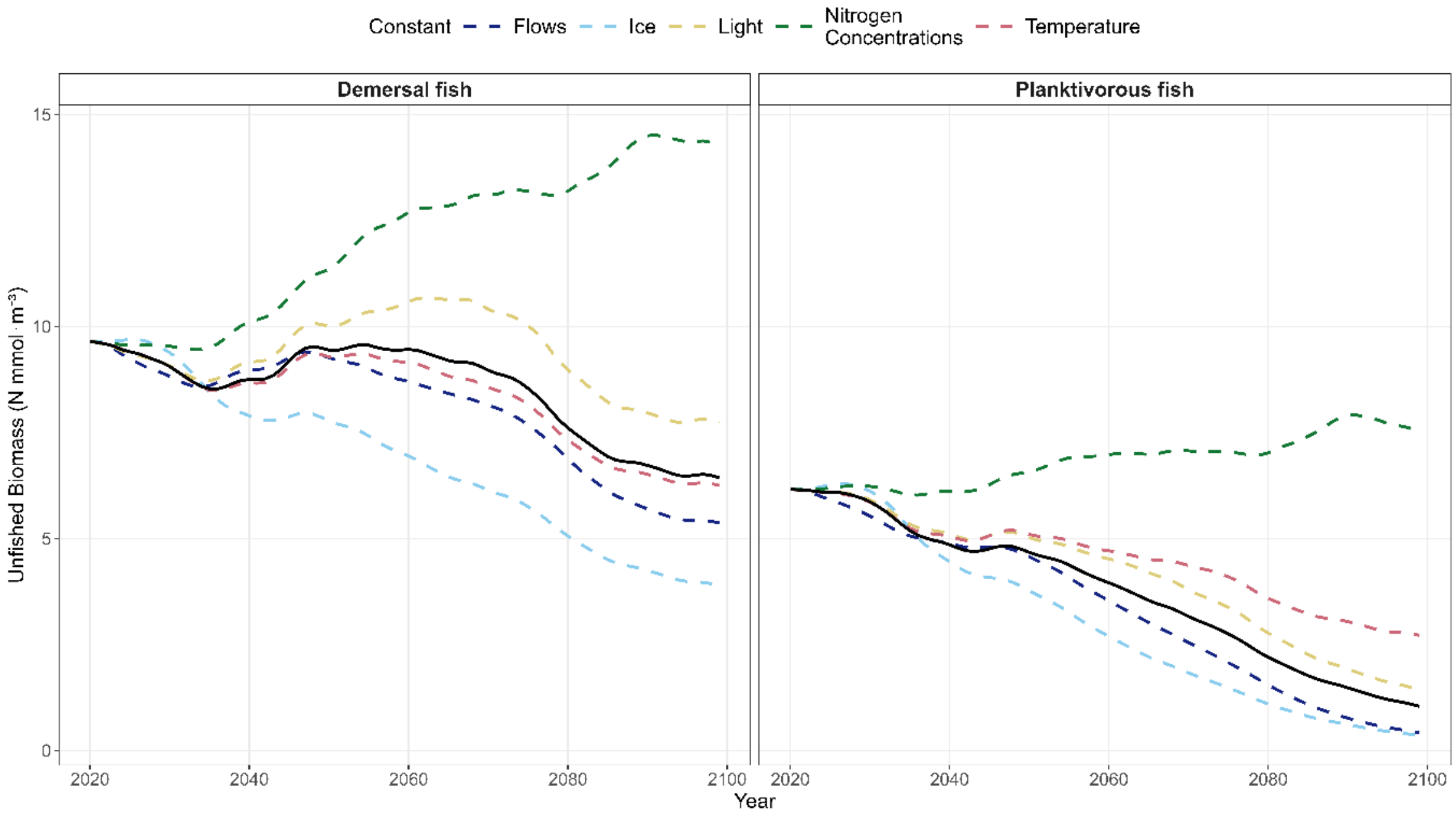

***Figure 6 Changes in Demersal and Planktivorous unfished biomass under a climate driver experiment.*** *Demersal and Planktivorous unfished biomass when different groups of climate drivers are held constant. Groupings are the same as those found in Table 2. Unfished biomass is tracked on a yearly basis, with individual groups held constant at their 2010-2020s smoothed values (dashed coloured lines). Per experiment, one group is held constant while the others are subject to climate change. When no groups are held constant, the unfished Demersal and Planktivorous biomass follows the solid black line.*

For both Demersal and Planktivorous fish, the Flows, Ice, Light, and Temperature driver groups all broadly follow the expected biomass trajectory (Figure 6), indicating that these drivers are not the causative features of change in the climate projection. Holding the Nitrogen Concentrations group constant presents a divergent trajectory for both guilds. For Demersal fish, unfished biomass increases from 9.65 to 14.28 $N\ mmol \cdot m^{-3}$ over the course of the century. Planktivorous fish also show an increase, though more modest, rising from 6.16 to 7.52 $N\ mmol \cdot m^{-3}$.

## 3.5 Ecosystem-Wide Recovery

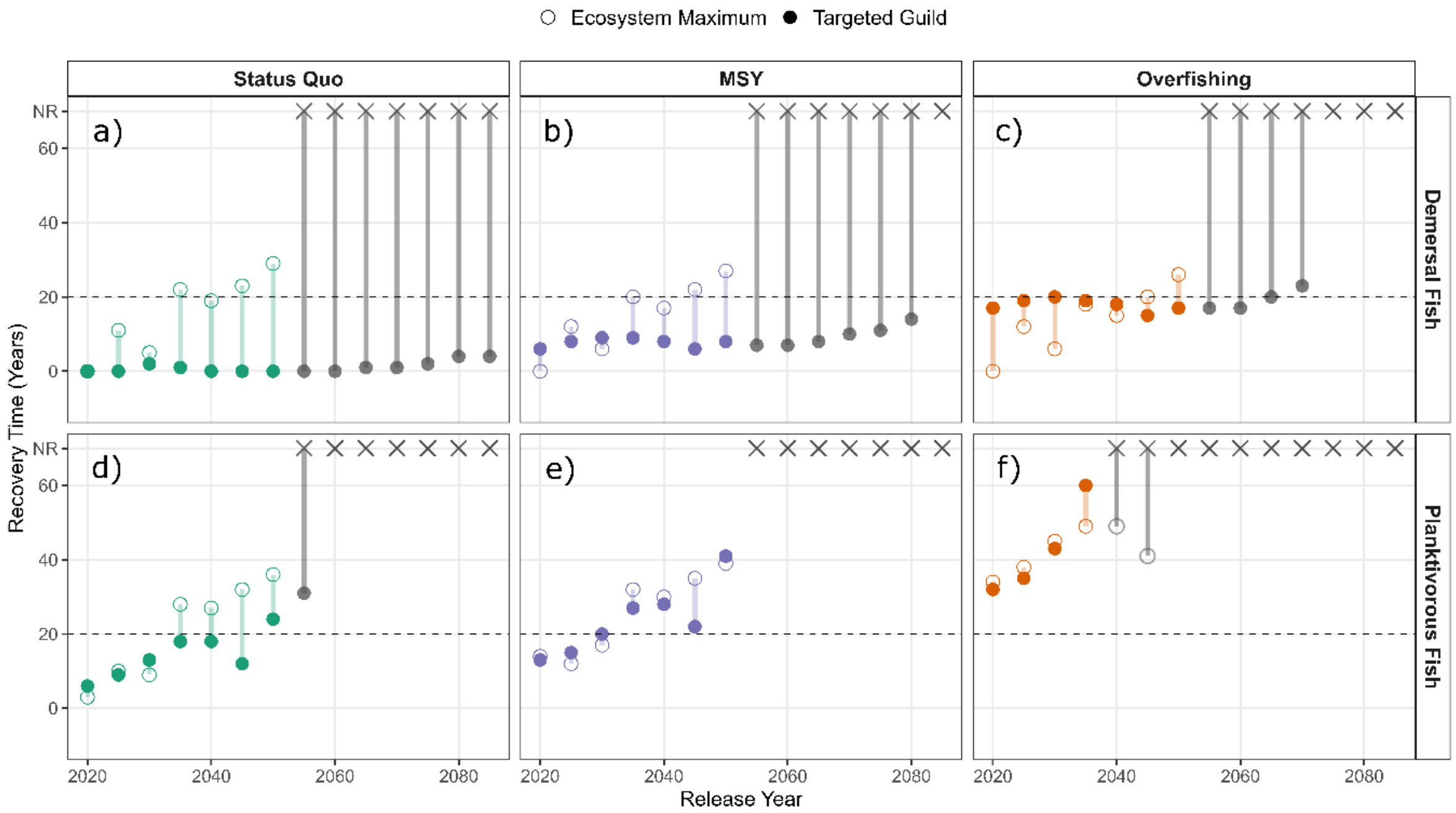

***Figure 7 Ecosystem-wide Recovery**. Ecosystem-wide recovery under Demersal (a, b, c) and Planktivorous (d, e, f) harvesting with the Status Quo (a, d), $MSY_{2020}$ (b, e), and Overfishing (c, f) scenarios. Labels for the longest to recover guild are presented within each panel. Filled circles represent the targeted guild recovery times, while open circles represent the maximum ecosystem-wide recovery time. Grey lines, marked with an 'X' at their maximum indicate release years in which either the targeted guild or the ecosystem maximum was not able to recover within the remaining simulation time. A marker for the 20-year MSC threshold is present within each panel (black dashed horizontal line). Recovery trajectories for each guild within StrathE2EPolar can be found in Supplementary Figures 5-15 (Demersal fishing) and 16-26 (Planktivorous fishing).*

### *3.5.1 Demersal fishing*

Under the status quo scenario (Figure 7a), ecosystem-wide recovery times increased from 0 years in 2020 to 22 years by 2035, exceeding the 20-year MSC threshold. Following this, recovery times continued to show increasing trends, to a peak of 29 years in 2050. Maritime mammals consistently exhibited the longest recovery times until 2050, after which recovery within the century became unachievable. Throughout the simulation period, ecosystem-wide recovery times were consistently longer than those of the targeted guild. A ranking of each guild's recovery time can be found in Supplementary Table 1.

Under the $MSY_{2020}$ scenario (Figure 7b) similar patterns emerged. Ecosystem-wide recovery remained largely slower than that of Demersal fish. Recovery peaked and exceeded the 20-year MSC threshold in 2050 at 27 years, driven by slow recovery in Maritime Mammals. Following this, recovery became impossible for the ecosystem before the end of the century.

In the Overfishing scenario (Figure 7c), ecosystem-wide recovery time generally increased, starting at 0 years in 2020 and rising to a peak of 26 years by 2050, once again driven by the slow recovery of the Maritime Mammals. Beyond 2050, ecosystem-wide recovery became impossible before the end of the century, despite Demersal fish recovery being comparatively fast. Approaching the mid-century, recovery times increasingly exceeded both the MSC's 20-year threshold and the recovery time of Demersal fish.

### *3.5.2 Planktivorous fishing*

Ecosystem-wide recovery under the Planktivorous fish disturbance scenarios diverged substantially from those observed under Demersal fish disturbances (Figure 7 d, e, f). Recovery times were highly sensitive to both climatic conditions and fishing pressure. Notably, the functional guild contributing the longest recovery time fluctuated amongst three higher trophic level guilds: Birds, Cetaceans, and Maritime Mammals. A ranking of each guild's recovery time can be found in Supplementary Table 2.

The Status Quo fishing scenario revealed that the guild associated with the longest ecosystem-wide recovery time was largely the Maritime Mammals (Figure 7d). Ecosystem-wide recovery time generally increased over the course of the century. Starting from a recovery time of 3 years in 2020, ecosystem-wide recovery showed a gradual upward trend, with recovery time surpassing the 20-year MSC threshold by 2035 and peaking at 36 years in 2050. Following this, ecosystem-wide recovery became impossible for the remainder of the century. Notably, ecosystem-wide recovery time exceeded that of Planktivorous fish for much of the early period, until 2060, when recovery became impossible for both the targeted guild and the whole ecosystem.

In the $MSY_{2020}$ fishing scenario, ecosystem-wide recovery in 2020 closely aligned with that of Planktivorous fish, at 14 and 13 years, respectively (Figure 7e). Recovery times then gradually increased, surpassing the 20-year MSC threshold by 2035 and remaining above it through 2040, 2045, and 2050, with recovery times of 30, 35, and 39 years. Beyond 2050, both ecosystem-wide and Planktivorous fish recovery became impossible before the end of the century. Throughout the century, the guild contributing the longest recovery time shifted from Maritime Mammals in 2020, to Birds in 2025, and Cetaceans in 2030. From 2035 to 2050, however, Maritime Mammals consistently represented the slowest recovering guild, peaking at 39 years.

The Overfishing scenario suggested that under elevated fishing pressure, Cetaceans consistently acted as the limiting guild, maximising recovery time between 2020 and 2040, peaking at 49 years (Figure 7e). Notably, in 2040, a shift occurred which was not present under the previous fishing scenarios. Ecosystem-wide recovery became faster than Planktivorous fish recovery due to recovery for the targeted guild becoming impossible. This was followed by a shift to a limiting guild of Maritime Mammals in 2045, beyond which recovery became impossible before the end of the century.

# 4 Discussion

Our analysis indicates that climate change alters the resilience of Demersal and Planktivorous fish to fishing pressure, though the strength of the effect is guild specific.

## 4.1 Demersal Fish Exhibit Resilience to Fishing Pressure Despite Increasing Recovery Times

Demersal fish biomass in the Barents Sea is projected to decline gradually over the course of the century, even in the absence of fishing pressure (Figure 3). In our simulations, this decline is primarily linked to changes in nitrogen concentrations at the model boundaries (Figure 6).

The Barents Sea is strongly influenced by the inflow of nutrient-rich Atlantic water, which can alter nutrient supply and subsequently primary production across the region (Noh et al. 2024). Declining nutrient concentrations, as a result of reduced inflow of Atlantic water (Årthun et al. 2019; Gerland et al. 2023), reduce primary production and limit food availability for Demersal species, leading to a bottom-up decline in Demersal biomass. A study by Geoffroy et al. (2023) suggests that some Demersal species, such as Cod, have a life cycle which is closely linked with sea-ice, suggesting vulnerability in the future due to a projected decrease in sea-ice concentration in the Barents Sea (Figure 1; Yool et al. 2013). Nevertheless, the cryospheric drivers were not as dominant a driver of Demersal fish biomass declines within our study.

Despite their relatively slow life histories (Bouchard et al. 2017; LeBlanc et al. 2020), Demersal fish maintained recovery times at or below the 20-year MSC threshold throughout most of the simulation (Figure 5). This rapid recovery in Demersal fish biomass supports the findings of Frank et al. (2011) and Liu et al. (2024), which suggest that Benthic species, a key prey guild for Demersal fish within our model, show strong resilience to large-scale ecosystem changes following major food web disruptions caused by overfishing. Notably, even under overfishing conditions, recovery to the 20-year MSC threshold appears feasible up to the mid-century when considering Demersal fish biomass in isolation, though this is still to a lower baseline biomass than was seen in 2020.

Adding to the work of Petrie et al. (2009), who found that warmer, species-rich southern fisheries tend to sustain trophic structure and consequently exhibit faster recovery under warming and moderate exploitation, our results indicate that recovery times for Demersal fish in the colder Barents Sea will progressively increase under future climate scenarios. While warming may enhance resilience in southern shelf seas by supporting richer, more productive communities, the Barents Sea is strongly influenced by declining nutrient inputs through Atlantic inflow (Figures 1 and 6; Noh et al. (2024)), which limits primary production and bottom-up energy transfer to the mid-trophic levels.

Consequently, despite warming, the reduced nutrient availability slows recovery and constrains growth.

It remains essential for fisheries managers to closely monitor changes in Demersal fish biomass in response to ongoing climatic changes, particularly given the projected rise in recovery time later in the century. Monitoring prey availability and diet composition alongside biomass trends is also important, as trophic interactions play an important role in the resilience of a system (Rose 2005; Ruzicka et al. 2024).

## 4.2 Planktivorous Fish Biomass and Recovery are Sensitive to Climate Change

Planktivorous fish biomass could decrease by as much as 80% by the end of the century, indicating a high sensitivity to climate change within our model. This finding aligns with our previous work using the same model (Heath et al. 2022), which projected a 15% reduction in Planktivorous fish biomass by the 2040s. The magnitude of the projected decline without any fishing activity suggests that climate change alone poses a considerable threat to the long-term abundance of Planktivorous stocks. As was the case with the Demersal fish, this decline is driven largely by a decrease in nitrogen concentrations at the model boundary (Figure 6), which is consistent with Årthun et al. (2019), who predicts a reduced inflow of nutrient-rich Atlantic water to the Barents Sea in the future.

A decline in dissolved inorganic nutrients (DIN; Figure 1) would reduce nutrient availability in the system, with cascading effects on plankton communities and, in turn, the prey base for Planktivorous fish (Drinkwater et al. 2021; Mueter et al. 2021). Despite the extensive work however, previous modelling studies remain uncertain as to whether primary productivity will increase, decrease, or remain stable over the coming decades (Slagstad et al. 2011; Skaret et al. 2014; Sandø et al. 2021; Gerland et al. 2023; Årthun et al. 2025). How primary productivity responds to changing nutrient conditions will directly influence the availability of prey for Planktivorous fish and, consequently, their role in supporting higher trophic levels.

Given this central role in the food web (Sivel et al. 2021), a decrease in prey options, and subsequently, Planktivorous fish resilience could have far-reaching implications for ecosystem-wide recovery and structuring (Ruzicka et al. 2024). Understanding the direction of future primary productivity in the Barents Sea is therefore vital due to its impact on secondary production (Søreide et al. 2010; Varpe 2012).

## 4.3 Ecosystem-wide recovery

When evaluating the recovery time of an ecosystem following a targeted disturbance, it is essential to consider not only the recovery of the directly impacted guild, but also the responses of non-target groups within the food web (Fulton et al. 2005; Perryman et al. 2021). In this study, rather than focusing solely on the biomass of the exploited guilds, we assessed the recovery time of all functional guilds within StrathE2EPolar to comprehensively assess ecosystem-wide recovery (Figure 5). It is important to note that in some cases, the unfished baseline could result in a lower biomass for present guilds than the fished scenario due to trophic cascades (for example, see Supplementary Figure 12 and 14).

#### 4.3.1 Ecosystem Limited Recovery Resistant to Demersal Harvesting, but Highly Responsive to Climate Change

Higher trophic level guilds, such as Birds, Cetaceans, and Maritime Mammals, exhibited the slowest recovery times (Figure 7a), consistent with work by Prato et al. (2013). This is due to a combination of life-history traits including longer reproductive cycles (Harding et al. 2009) and lower reproductive rates (Deb and Bailey 2023), which in our model are reflected by density-dependent prey uptake functions that restrict population growth even when prey are abundant, thereby reproducing the slow recovery characteristic of these life-history traits.

Ecosystem-wide recovery times showed an increasing trend until 2050, highlighting increased stress on the ecosystem due to climate change (van Nes and Scheffer 2007). However, we observed that ecosystem-wide recovery was relatively insensitive to changes in Demersal fishing pressure. Only in the most extreme cases, the overfishing scenario, would the Demersal recovery time exceed that of the whole ecosystem. This suggests that the collapse of the Demersal guild had limited knock-on effects on the wider ecosystem, due to high trophic levels having a low feeding preference for Demersal fish within our model. This finding is in line with that of Aune et al. (2018), who found that Demersal species have a high redundancy in the Barents Sea. Our model suggests that harvesting Demersal fish is less likely to trigger large-scale ecosystem collapse in the Barents Sea. However, this possibility would have to be very carefully explored by fisheries managers using species specific models designed for tactical management decisions.

Notably, across all three Demersal fishing pressure scenarios, the model projected that recovery to the dynamic MSC threshold by the end of the century would become infeasible in 2050. In particular, Maritime Mammals (Polar Bears) consistently showed the longest recovery among all guilds.

The progressive loss of suitable habitat for Maritime Mammals plays a major role. These ice-dependent species are facing the erosion of their habitat as sea ice continues to retreat (Regehr et al. 2016). Projections indicate that the Barents Sea may become seasonally ice-free by the mid-century (Bonan et al. 2021; Rieke et al. 2023), leaving these species without a viable environment in which to live and hunt (Chen 2022). As their habitat disappears, so too does their biomass within the system, leading to local, and potentially global (Notz and Community 2020; Årthun et al. 2021), extinction due to climate alone.

The broader effects of climate change on marine food web structure amplify the decline in suitable predator habitat. Ullah et al. (2018) found that future climate change can weaken marine food webs by reducing the transfer of energy to higher trophic levels, thereby diminishing the biomass of top predators, such as Maritime Mammals, causing current recovery targets to become unattainable.

### 4.3.2 Planktivorous fish Harvesting Coupled with Climate Change Leads to an Increase in Ecosystem-Wide Recovery

Consistent with patterns observed under Demersal harvesting, higher trophic level guilds (Birds, Cetaceans, and Maritime Mammals) also exhibited the slowest recovery rates under the Planktivorous fishing scenarios (Figure 7d, e, f). However, unlike the Demersal case, there was less consistency in which guild exhibited the slowest recovery. For example, within the status quo scenario, Birds and Maritime Mammals alternated as the slowest guild to recover between 2020 and 2035. In line with the findings of Sivel et al. (2021), these shifts suggest that Planktivorous fish play a key role as a mid-trophic level species in the Barents Sea. Our findings reemphasise the work of Essington et al. (2015), who suggests that Planktivorous fish harvesting, coupled with climate effects, may lead to collapses in higher trophic level biomass.

Ecosystem-wide recovery times exhibited a consistent upward trend across all three Planktivorous fishing scenarios, suggesting a general decline in ecosystem resilience over the century. Recovery times exhibited sensitivity to fishing disturbances across all years. In particularly, the sensitivity which is present in the early century (2020-2030) could show the immediate threat Planktivorous fishing may have on the wider ecosystem and justifies an ecosystem-wide approach to the management of the stock (Garcia 2003; Heath et al. 2022).

These results highlight the vulnerability of the Barents Sea ecosystem to changes in Planktivorous fish biomass (Ruzicka et al. 2024) and emphasise the need for climate action and ecosystem-based fisheries management (Garcia 2003).

## 4.4 Strengths and Limitations

#### *Model Limitations*

While the time-varying physical and chemical drivers (e.g. Seawater Temperature, Ice Cover, Nutrient Concentrations) are updated according to climate model projections when running in transient mode, other aspects of the system (e.g. Fishing Distribution and Event Timings) are held constant in each interval year, despite our previous work indicating that they may also change (Hatton et al. 2025). Consequently, our results are conditional on the assumption that some input parameters will remain constant into the future.

Although the causative drivers of unfished biomass were explored (Section 2.5), it should be noted that this experiment only accounted for direct effects. The Temperature grouping (Table 2), for example, only explores the physiological effects of Temperature, but not the effect of stratification and mixing. It has been suggested by Gerland et al. (2023) and Noh et al. (2024) that an increase in Sea Water Temperature will enhance productivity through reduced sea-ice and weakened stratification in the future, however this was not explored within this study.

### *Recovery Criteria*

Defining recovery relative to an absolute biomass target assumes that ecosystem capacity is stationary, allowing for historical biomasses to be the target for recovery. This study opted to base recovery on a dynamically calculated biomass set out by the MSC (Council 2022), offering an approach which calculated new limits on the target guild biomasses for the Barents Sea. Both approaches carry trade-offs. Defining recovery based on a fixed biomass target may become unrealistic as climate change has been shown to substantially alter Arctic ecosystems (Aschan et al. 2013; Heath et al. 2022; Laverick et al. 2025). In such a system, aiming for past reference points risks setting unachievable goals that divert attention from managing the ecosystem within its new and evolving state. On the other hand, using a shifting unfished baseline as the reference point may undermine faith in MSC certification, as recovery to a lower target may not satisfy stakeholders. Future management will need to consider these two views on reference points to ensure recovery criteria remain meaningful in a rapidly changing environment.

### *Ecosystem Indicators*

Our approach offers a comprehensive and interpretable measure of resilience by directly tracking change across all components of the ecosystem, rather than condensing ecosystem dynamics into a single aggregated metric. While this study uses recovery time and biomass to assess ecosystem-wide recovery, it is important to recognise that a range of ecological indicators could provide complementary perspectives. Alternative approaches, such as network-based food web indicators, may yield additional insights. Metrics derived from ecological network analysis, particularly those developed by Ulanowicz (1986), could reveal aspects of resilience specific to certain stocks within the Barents Sea ecosystem. For example, it has been suggested by Rajpar et al. (2018) that Birds may be a useful indicator for the overall health of an ecosystem due to their trophic position and ability to feed down the food web. Whereas these indicators condense ecosystem dynamics into a single value, our approach examines recovery times for all guilds, with the longest recovery time taken as representative of the overall ecosystem's resilience.

# 5 Conclusion

This study identifies the contrasting resilience of Demersal and Planktivorous fish in the Barents Sea under future climate change and fishing scenarios.

***How do combined climate alterations and fishing pressures affect the recovery trajectories, timescales, and resilience of Demersal and Planktivorous fish stocks in the Barents Sea?***

The recovery times of both Demersal and Planktivorous fish biomass are set to increase due to climate shifts. An Increase in harvesting pressure would still lead to Demersal fish biomass recovery within the 20-year MSC threshold, suggesting that Demersal fish are relatively resilient to climate and harvesting stress, although recovery times will lengthen in the future. Climate effects would force some high trophic level guilds, such as Birds, Cetaceans, and Maritime Mammals outside the MSC recovery window, indicating harm to the ecosystem and a reduced capacity to support fisheries. Planktivorous fish have high recovery times and are both sensitive to climate and harvesting pressures. These findings suggest that stocks must be carefully managed in the future as resilience is expected to decline further. Adding to this, the harvesting of Planktivorous fish could have very serious consequences for the wider ecosystem, specifically higher trophic levels.

***How do ecosystem-wide recovery times compare to those of target fish stocks, and what does this reveal about the effectiveness of single-guild recovery as a proxy for ecosystem health?***

Demersal fish biomass recovery times mostly remained below the 20-year MSC threshold for all three fishing scenarios. However, due to shifts in climate, ecosystem-wide recovery is set to increase beyond this threshold, suggesting that single-guild recovery would be a poor proxy for ecosystem health.

In the case of the Planktivorous fish, ecosystem-wide recovery time consistently exceeded the 20-year MSC threshold for all three fishing scenarios. In the near future, focussing solely on the recovery of Planktivorous fish would have negative consequences for the wider ecosystem. In this case, a focus on single-guild recovery as a proxy for ecosystem health cannot be recommended.

***How should recovery criteria be defined under future climate and harvest conditions?***

The Marine Stewardship Council's 20-year, 80% recovery threshold performs reasonably well under present and near-future conditions for more resilient stocks, such as the Demersal fish in this study. Under severe climate change, and for sensitive guilds such as Planktivorous fish, achieving the benchmark of 80% recovery within 20 years is particularly challenging due to reductions in the unfished biomass throughout the course of the century. This highlights a wider challenge in defining recovery criteria. Static biomass targets assume that ecosystem capacity is constant, yet in a rapidly changing Arctic this assumption is unlikely to hold. In systems experiencing long-term climate effects, it is therefore important to complement historical recovery targets with dynamic thresholds that account for the changing ecosystem, recognising that shifting baselines challenge the effectiveness of static targets. At the same time, monitoring the biomasses of higher trophic level guilds remains vital to ensure ecosystem health, while also considering absolute thresholds where critical ecosystem functions or services must be maintained.

Alleway, Heidi K., Emily S. Klein, Liz Cameron, et al. 2023. 'The Shifting Baseline Syndrome as a Connective Concept for More Informed and Just Responses to Global Environmental Change'. *People and Nature* 5 (3): 885–96. https://doi.org/10.1002/pan3.10473.

Arctic Council. 2015. *ENHANCED BLACK CARBON AND METHANE EMISSIONS REDUCTIONS AN ARCTIC COUNCIL FRAMEWORK FOR ACTION.* http://hdl.handle.net/11374/610.

Årthun, Marius, Khuong V. Dinh, Jakob Dörr, et al. 2025. 'The Future Barents Sea—A Synthesis of Physical, Biogeochemical, and Ecological Changes toward 2050 and 2100'. *Elem Sci Anth* 13 (1): 00046. https://doi.org/10.1525/elementa.2024.00046.

Årthun, Marius, Tor Eldevik, and Lars H. Smedsrud. 2019. *The Role of Atlantic Heat Transport in Future Arctic Winter Sea Ice Loss*. Journal of Climate. June 1. https://doi.org/10.1175/JCLI-D-18-0750.1.

Årthun, Marius, Ingrid H. Onarheim, Jakob Dörr, and Tor Eldevik. 2021. 'The Seasonal and Regional Transition to an Ice-Free Arctic'. *Geophysical Research Letters* 48 (1): e2020GL090825. https://doi.org/10.1029/2020GL090825.

Aschan, Michaela, Maria Fossheim, Michael Greenacre, and Raul Primicerio. 2013. 'Change in Fish Community Structure in the Barents Sea'. *PLoS ONE* 8 (4): 4. https://doi.org/10.1371/journal.pone.0062748.

Aune, Magnus, Michaela M. Aschan, Michael Greenacre, Andrey V. Dolgov, Maria Fossheim, and Raul Primicerio. 2018. 'Functional Roles and Redundancy of Demersal Barents Sea Fish: Ecological Implications of Environmental Change'. *PLoS ONE* 13 (11): e0207451. https://doi.org/10.1371/journal.pone.0207451.

Bogstad, Bjarte, Harald GjÃ¸sÃ¦ter, Tore Haug, and Ulf LindstrÃ¸m. 2015. 'A Review of the Battle for Food in the Barents Sea: Cod vs. Marine Mammals'. *Frontiers in Ecology and Evolution* 3 (March). https://doi.org/10.3389/fevo.2015.00029.

Bonan, David B., Tapio Schneider, Ian Eisenman, and Robert C. J. Wills. 2021. 'Constraining the Date of a Seasonally Ice-Free Arctic Using a Simple Model'. *Geophysical Research Letters* 48 (18): e2021GL094309. https://doi.org/10.1029/2021GL094309.

Bouchard, Caroline, Maxime Geoffroy, Mathieu LeBlanc, et al. 2017. 'Climate Warming Enhances Polar Cod Recruitment, at Least Transiently'. *Progress in Oceanography* 156 (August): 121–29. https://doi.org/10.1016/j.pocean.2017.06.008.

Callaghan, Terry V., Lars Olof Björn, Yuri Chernov, et al. 2004. 'Biodiversity, Distributions and Adaptations of Arctic Species in the Context of Environmental Change'. *AMBIO: A Journal of the Human Environment* 33 (7): 404–17. https://doi.org/10.1579/0044-7447-33.7.404.

Chan, Farrah T., Keara Stanislawczyk, Anna C. Sneekes, et al. 2019. 'Climate Change Opens New Frontiers for Marine Species in the Arctic: Current Trends and Future Invasion Risks'. *Global Change Biology* 25 (1): 25–38. https://doi.org/10.1111/gcb.14469.

Chavas, Jean-Paul, and Corbett Grainger. 2019. 'On the Dynamic Instability of Arctic Sea Ice'. *Npj Climate and Atmospheric Science* 2 (1): 23. https://doi.org/10.1038/s41612-019-0080-x.

Chen, Jasmin. 2022. 'Impacts of Sea Ice Loss on Polar Bear Diet, Prey Availability, Foraging Behaviors, and Human-Bear Interactions in the Arctic'. *Master's Projects and Capstones*, May 21. https://repository.usfca.edu/capstone/1350.

Cochrane, Kevern, Cassandra De Young, Doris Soto, and Tarûb Bahri. 2009. 'Climate Change Implications for Fisheries and Aquaculture'. *FAO Fisheries and Aquaculture Technical Paper* 530: 212.

Council, Marine Stewardship. 2022. 'MSC Fisheries Standard'. *Version* 3: 26.

Deb, Jiban Chandra, and Sarah A. Bailey. 2023. 'Arctic Marine Ecosystems Face Increasing Climate Stress'. *Environmental Reviews* 31 (3): 403–51. https://doi.org/10.1139/er-2022-0101.

Dolgov, AV. 2016. 'Composition, Formation and Trophic Structure of the Barents Sea Fish Communities'. *PINRO, Murmansk (In Russian)*, 336.

Drinkwater, Kenneth F, Naomi Harada, Shigeto Nishino, et al. 2021. 'Possible Future Scenarios for Two Major Arctic Gateways Connecting Subarctic and Arctic Marine Systems: I. Climate and Physical–Chemical Oceanography'. *ICES Journal of Marine Science* 78 (9): 3046–65. https://doi.org/10.1093/icesjms/fsab182.
Emblemsvåg, Margrete, Karl Michael Werner, Ismael Núñez-Riboni, et al. 2022. 'Deep Demersal Fish Communities Respond Rapidly to Warming in a Frontal Region between Arctic and Atlantic Waters'. *Global Change Biology* 28 (9): 2979–90. https://doi.org/10.1111/gcb.16113.
Eriksen, E., B. Husson, G. Skaret, et al. 2025. 'The Living Barents Sea Response to Peak-Warming and Subsequent Cooling'. *Scientific Reports* 15 (1): 13008. https://doi.org/10.1038/s41598-025-96964-x.
Essington, Timothy E., Pamela E. Moriarty, Halley E. Froehlich, et al. 2015. 'Fishing Amplifies Forage Fish Population Collapses'. *Proceedings of the National Academy of Sciences* 112 (21): 6648–52. https://doi.org/10.1073/pnas.1422020112.
Fall, Johanna, Lorenzo Ciannelli, Georg Skaret, and Edda Johannesen. 2018. 'Seasonal Dynamics of Spatial Distributions and Overlap between Northeast Arctic Cod (Gadus Morhua) and Capelin (Mallotus Villosus) in the Barents Sea'. *PLOS ONE* 13 (10): e0205921. https://doi.org/10.1371/journal.pone.0205921.
Frank, Kenneth T., Brian Petrie, Jae S. Choi, and William C. Leggett. 2005. 'Trophic Cascades in a Formerly Cod-Dominated Ecosystem'. *Science* 308 (5728): 1621–23. https://doi.org/10.1126/science.1113075.
Frank, Kenneth T., Brian Petrie, Jonathan A. D. Fisher, and William C. Leggett. 2011. 'Transient Dynamics of an Altered Large Marine Ecosystem'. *Nature* 477 (7362): 86–89. https://doi.org/10.1038/nature10285.
Fulton, E, A Smith, and A Punt. 2005. 'Which Ecological Indicators Can Robustly Detect Effects of Fishing?' *ICES Journal of Marine Science* 62 (3): 540–51. https://doi.org/10.1016/j.icesjms.2004.12.012.
Garcia, S. M. 2003. *The Ecosystem Approach to Fisheries: Issues, Terminology, Principles, Institutional Foundations, Implementation and Outlook*. Food & Agriculture Org.
Geoffroy, Maxime, Caroline Bouchard, Hauke Flores, et al. 2023. 'The Circumpolar Impacts of Climate Change and Anthropogenic Stressors on Arctic Cod ( *Boreogadus Saida* ) and Its Ecosystem'. *Elem Sci Anth* 11 (1): 00097. https://doi.org/10.1525/elementa.2022.00097.
Gerland, Sebastian, Randi B. Ingvaldsen, Marit Reigstad, et al. 2023. 'Still Arctic?—The Changing Barents Sea'. *Elem Sci Anth* 11 (1): 00088. https://doi.org/10.1525/elementa.2022.00088.
Gjøsæter, Harald. 1998. 'The Population Biology and Exploitation of Capelin ( *Mallotus Villosus* ) in the Barents Sea'. *Sarsia* 83 (6): 453–96. https://doi.org/10.1080/00364827.1998.10420445.
Gjøsæter, Harald, Sigurd Tjelmeland, and Bjarte Bogstad. 2012. 'Ecosystem-Based Management of Fish Species in the Barents Sea'. *Global Progress in Ecosystem-Based Fisheries Management*, 333–52.
H, Wickham. 2015. 'Dplyr: A Grammar of Data Manipulation.' In *R Package Version 04.*, vol. 3. https://cir.nii.ac.jp/crid/1370584339779773824.
Hansen, Cecilie, Solfrid Sætre Hjøllo, Morten D Skogen, et al. 2024. 'The Combined Effects of Warming, Ocean Acidification, and Fishing on the Northeast Atlantic Cod ( *Gadus Morhua* ) in the Barents Sea'. *ICES Journal of Marine Science* 81 (5): 877–86. https://doi.org/10.1093/icesjms/fsae042.
Harding, Ann M.A., Alexander S. Kitaysky, Margaret E. Hall, et al. 2009. 'Flexibility in the Parental Effort of an Arctic-Breeding Seabird'. *Functional Ecology* 23 (2): 348–58. https://doi.org/10.1111/j.1365-2435.2008.01488.x.
Hatton, Matthew, Jack Laverick, Neil Banas, Elliot Sivel, and Michael Heath. 2025. 'Climate Change Opens Up New Fishing Possibilities for Large-Scale Trawling Vessels Off West Greenland'. *Fisheries Oceanography*, May 12, e12736. https://doi.org/10.1111/fog.12736.
Heath, Michael R., Déborah Benkort, Andrew S. Brierley, et al. 2022. 'Ecosystem Approach to Harvesting in the Arctic: Walking the Tightrope between Exploitation and Conservation in the Barents Sea'. *Ambio* 51 (2): 456–70. https://doi.org/10.1007/s13280-021-01616-9.

Hobday, Alistair J. 2015. ‘Shifting Baselines: The Past and the Future of Ocean Fisheries’. *Oceanography* 25 (1): 306–8. https://doi.org/10.5670/oceanog.2012.35.

Holling, C. S. 1973. ‘Resilience and Stability of Ecological Systems’. *Annual Review of Ecology and Systematics* 4: 1–23. JSTOR.

Huse, Geir, and Ingrid Ellingsen. 2008. ‘Capelin Migrations and Climate Change – a Modelling Analysis’. *Climatic Change* 87 (1–2): 177–97. https://doi.org/10.1007/s10584-007-9347-z.

Huserbråten, Mats Brockstedt Olsen, Elena Eriksen, Harald Gjøsæter, and Frode Vikebø. 2019. ‘Polar Cod in Jeopardy under the Retreating Arctic Sea Ice’. *Communications Biology* 2 (1): 407. https://doi.org/10.1038/s42003-019-0649-2.

Hvingel, Carsten, and Fabian Zimmermann. 2023. ‘Advice on Fishing Opportunities for Barents Sea Shrimp in 2024’. *IMR/PINRO Joint Report Series*.

ICES. 2021. *Barents Sea Ecoregion Fisheries Overview*. https://doi.org/10.17895/ICES.ADVICE.9166.

Ingvaldsen, Randi B., Karen M. Assmann, Raul Primicerio, Maria Fossheim, Igor V. Polyakov, and Andrey V. Dolgov. 2021. ‘Physical Manifestations and Ecological Implications of Arctic Atlantification’. *Nature Reviews Earth & Environment* 2 (12): 874–89. https://doi.org/10.1038/s43017-021-00228-x.

Johannesen, Edda, Randi B. Ingvaldsen, Bjarte Bogstad, et al. 2012. ‘Changes in Barents Sea Ecosystem State, 1970–2009: Climate Fluctuations, Human Impact, and Trophic Interactions’. *ICES Journal of Marine Science* 69 (5): 880–89. https://doi.org/10.1093/icesjms/fss046.

Jørgensen, Lis L., Elizabeth A. Logerwell, Natalia Strelkova, et al. 2022. ‘International Megabenthic Long-Term Monitoring of a Changing Arctic Ecosystem: Baseline Results’. *Progress in Oceanography* 200 (January): 102712. https://doi.org/10.1016/j.pocean.2021.102712.

Kaartvedt, Stein, and Josefin Titelman. 2018. ‘Planktivorous Fish in a Future Arctic Ocean of Changing Ice and Unchanged Photoperiod’. *ICES Journal of Marine Science* 75 (7): 2312–18. https://doi.org/10.1093/icesjms/fsx248.

Kayode-Edwards, Jesudunni Otinu, Ifeoluwa Ihotu Kayode-Edwards, and Damilola Olohi Kayode-Edwards. 2024. ‘Climate Change in the Arctic’. In *Arctic Marine Ecotoxicology*, edited by Patrick Omoregie Isibor. Springer Nature Switzerland. https://doi.org/10.1007/978-3-031-73584-4_4.

Kortsch, Susanne, Raul Primicerio, Maria Fossheim, Andrey V. Dolgov, and Michaela Aschan. 2015. ‘Climate Change Alters the Structure of Arctic Marine Food Webs Due to Poleward Shifts of Boreal Generalists’. *Proceedings of the Royal Society B: Biological Sciences* 282 (1814): 20151546. https://doi.org/10.1098/rspb.2015.1546.

Laverick, Jack H., Douglas C. Speirs, and Michael R. Heath. 2025. ‘Sea-Ice Retreat From the Northeast Greenland Continental Shelf Triggers a Marine Trophic Cascade’. *Global Change Biology* 31 (4): e70189. https://doi.org/10.1111/gcb.70189.

LeBlanc, Mathieu, Maxime Geoffroy, Caroline Bouchard, et al. 2020. ‘Pelagic Production and the Recruitment of Juvenile Polar Cod Boreogadus Saida in Canadian Arctic Seas’. *Polar Biology* 43 (8): 1043–54. https://doi.org/10.1007/s00300-019-02565-6.

Liu, Guankui, Peng Sun, Jin Gao, Fabian Zimmermann, Yongjun Tian, and Mikko Heino. 2024. ‘Pelagic and Demersal Fish Population Rebuilding in Response to Fisheries-Induced Evolution in Exploited China Seas’. *Ecological Indicators* 168 (November): 112742. https://doi.org/10.1016/j.ecolind.2024.112742.

Lynam, Christopher P., Laura Uusitalo, Joana Patrício, et al. 2016. ‘Uses of Innovative Modeling Tools within the Implementation of the Marine Strategy Framework Directive’. *Frontiers in Marine Science* 3 (September). https://doi.org/10.3389/fmars.2016.00182.

Marine Stewardship Council. 2023. *MSC Fisheries Report - Barents Sea Cod, Haddock, and Saithe*. https://fisheries.msc.org/en/fisheries/barents-sea-cod-haddock-and-saithe/.

Marz, Stacey. 2010. *Arctic Sea Ice Ecosystem: A Summary of Species That Depend on and Associate with Sea Ice and Projected Impacts from Sea Ice Changes*.

Michael Heath, Robert Wilson and Douglas Speirs. 2015. *Modelling the Whole-Ecosystem Impacts of Trawling*. Fisheries Innovation Scotland (FIS).

Mueter, Franz J, Benjamin Planque, George L Hunt Jr, et al. 2021. 'Possible Future Scenarios in the Gateways to the Arctic for Subarctic and Arctic Marine Systems: II. Prey Resources, Food Webs, Fish, and Fisheries'. *ICES Journal of Marine Science* 78 (9): 3017–45. https://doi.org/10.1093/icesjms/fsab122.

Nes, Egbert H. van, and Marten Scheffer. 2007. 'Slow Recovery from Perturbations as a Generic Indicator of a Nearby Catastrophic Shift.' *The American Naturalist* 169 (6): 738–47. https://doi.org/10.1086/516845.

Noh, Kyung-Min, Ji-Hoon Oh, Hyung-Gyu Lim, Hajoon Song, and Jong-Seong Kug. 2024. 'Role of Atlantification in Enhanced Primary Productivity in the Barents Sea'. *Earth's Future* 12 (1): e2023EF003709. https://doi.org/10.1029/2023EF003709.

Notz, Dirk, and Simip Community. 2020. 'Arctic Sea Ice in CMIP6'. *Geophysical Research Letters* 47 (10): e2019GL086749. https://doi.org/10.1029/2019GL086749.

Olsen, Erik, Sondre Aanes, Sigbjørn Mehl, Jens Christian Holst, Asgeir Aglen, and Harald Gjøsæter. 2010. 'Cod, Haddock, Saithe, Herring, and Capelin in the Barents Sea and Adjacent Waters: A Review of the Biological Value of the Area'. *ICES Journal of Marine Science* 67 (1): 87–101. https://doi.org/10.1093/icesjms/fsp229.

Pauly, Daniel. 1995. 'Anecdotes and the Shifting Baseline Syndrome of Fisheries'. *Trends in Ecology & Evolution* 10 (10): 430. https://doi.org/10.1016/S0169-5347(00)89171-5.

Pebesma, Edzer. 2018. 'Simple Features for R: Standardized Support for Spatial Vector Data'. *The R Journal* 10 (1): 439. https://doi.org/10.32614/RJ-2018-009.

Perryman, H. A., C. Hansen, D. Howell, and E. Olsen. 2021. 'A Review of Applications Evaluating Fisheries Management Scenarios through Marine Ecosystem Models'. *Reviews in Fisheries Science & Aquaculture* 29 (4): 800–835. https://doi.org/10.1080/23308249.2021.1884642.

Petrie, Brian, Kenneth T. Frank, Nancy L. Shackell, and William C. Leggett. 2009. 'Structure and Stability in Exploited Marine Fish Communities: Quantifying Critical Transitions'. *Fisheries Oceanography* 18 (2): 83–101. https://doi.org/10.1111/j.1365-2419.2009.00500.x.

Prato, Giulia, Paolo Guidetti, Fabrizio Bartolini, Luisa Mangialajo, and Patrice Francour. 2013. 'The Importance of High-Level Predators in Marine Protected Area Management: Consequences of Their Decline and Their Potential Recovery in the Mediterranean Context'. *Advances in Oceanography and Limnology* 4 (2): 176–93. https://doi.org/10.1080/19475721.2013.841754.

Rajpar, Muhammad Nawaz, Ibrahim Ozdemir, Mohamed Zakaria, Shazia Sheryar, and Abdu Rab. 2018. 'Seabirds as Bioindicators of Marine Ecosystems'. In *Seabirds*, edited by Heimo Mikkola. InTech. https://doi.org/10.5772/intechopen.75458.

Regehr, Eric V., Kristin L. Laidre, H. Resit Akçakaya, et al. 2016. 'Conservation Status of Polar Bears ( *Ursus Maritimus* ) in Relation to Projected Sea-Ice Declines'. *Biology Letters* 12 (12): 20160556. https://doi.org/10.1098/rsbl.2016.0556.

Reygondeau, Gabriel, and Grégory Beaugrand. 2011. 'Future Climate-driven Shifts in Distribution of *Calanus Finmarchicus*'. *Global Change Biology* 17 (2): 756–66. https://doi.org/10.1111/j.1365-2486.2010.02310.x.

Riahi, Keywan, Shilpa Rao, Volker Krey, et al. 2011. 'RCP 8.5—A Scenario of Comparatively High Greenhouse Gas Emissions'. *Climatic Change* 109 (1–2): 33–57. https://doi.org/10.1007/s10584-011-0149-y.

Richardson, Anthony J. 2008. 'In Hot Water: Zooplankton and Climate Change'. *ICES Journal of Marine Science* 65 (3): 279–95. https://doi.org/10.1093/icesjms/fsn028.

Rieke, Ole, Marius Årthun, and Jakob Simon Dörr. 2023. 'Rapid Sea Ice Changes in the Future Barents Sea'. *The Cryosphere* 17 (4): 1445–56. https://doi.org/10.5194/tc-17-1445-2023.

Ripple, William J., James A. Estes, Oswald J. Schmitz, et al. 2016. 'What Is a Trophic Cascade?' *Trends in Ecology & Evolution* 31 (11): 842–49. https://doi.org/10.1016/j.tree.2016.08.010.

Ritchie, Michelle, Tim Frazier, Harley Johansen, and Erik Wood. 2021. 'Early Climate Change Indicators in the Arctic: A Geographical Perspective'. *Applied Geography* 135 (October): 102562. https://doi.org/10.1016/j.apgeog.2021.102562.

Roderfeld, Hedwig, Eleanor Blyth, Rutger Dankers, et al. 2008. 'Potential Impact of Climate Change on Ecosystems of the Barents Sea Region'. *Climatic Change* 87 (1–2): 283–303. https://doi.org/10.1007/s10584-007-9350-4.

Rose, G.A. 2005. 'Capelin (Mallotus Villosus) Distribution and Climate: A Sea "Canary" for Marine Ecosystem Change'. *ICES Journal of Marine Science* 62 (7): 1524–30. https://doi.org/10.1016/j.icesjms.2005.05.008.

Ruzicka, J, L Chiaverano, M Coll, et al. 2024. 'The Role of Small Pelagic Fish in Diverse Ecosystems: Knowledge Gleaned from Food-Web Models'. *Marine Ecology Progress Series* 741 (July): 7–27. https://doi.org/10.3354/meps14513.

Sandø, Anne Britt, Erik Askov Mousing, W P Budgell, Solfrid S Hjøllo, Morten D Skogen, and B Ådlandsvik. 2021. 'Barents Sea Plankton Production and Controlling Factors in a Fluctuating Climate'. *ICES Journal of Marine Science* 78 (6): 1999–2016. https://doi.org/10.1093/icesjms/fsab067.

Sivel, Elliot, Benjamin Planque, Ulf Lindstrøm, and Nigel G. Yoccoz. 2021. 'Multiple Configurations and Fluctuating Trophic Control in the Barents Sea Food-Web'. *PLOS ONE* 16 (7): e0254015. https://doi.org/10.1371/journal.pone.0254015.

Skaret, G., P. Dalpadado, S. S. Hjøllo, M. D. Skogen, and E. Strand. 2014. '*Calanus Finmarchicus* Abundance, Production and Population Dynamics in the Barents Sea in a Future Climate'. *Progress in Oceanography* 125 (June): 26–39. https://doi.org/10.1016/j.pocean.2014.04.008.

Slagstad, D., I. H. Ellingsen, and P. Wassmann. 2011. 'Evaluating Primary and Secondary Production in an Arctic Ocean Void of Summer Sea Ice: An Experimental Simulation Approach'. *Progress in Oceanography*, Arctic Marine Ecosystems in an Era of Rapid Climate Change, vol. 90 (1): 117–31. https://doi.org/10.1016/j.pocean.2011.02.009.

Smith, Kimberley A, Brett Crisafulli, Amber Quinn, Bianca Brooks, Gabby E Mitsopoulos, and Daniel Yeoh. 2025. *Fisheries Research Report No. 351: Ecological Risk Assessment for the Western Australian Southern Nearshore Fish Resource*.

Søreide, Janne E., Eva Leu, Jørgen Berge, Martin Graeve, and Stig Falk-Petersen. 2010. 'Timing of Blooms, Algal Food Quality and Calanus Glacialis Reproduction and Growth in a Changing Arctic'. *Global Change Biology* 16 (11): 3154–63. https://doi.org/10.1111/j.1365-2486.2010.02175.x.

Stokke, Olav Schram, Lee G Anderson, and Natalia Mirovitskaya. 1999. 'The Barents Sea Fisheries'. *The Effectiveness of International Environmental Regimes: Causal Connections and Behavioral Mechanisms*, 91–154.

Team, RDC. 2010. 'R: A Language and Environment for Statistical Computing'. https://cir.nii.ac.jp/crid/1370294721063650048.

Team, RStudio. 2020. *RStudio: Integrated Development Environment for R. RStudio, PBC*.

Thorpe, Robert B., Nina L. Arroyo, Georges Safi, et al. 2022. 'The Response of North Sea Ecosystem Functional Groups to Warming and Changes in Fishing'. *Frontiers in Marine Science* 9 (April). https://doi.org/10.3389/fmars.2022.841909.

Thorpe, Robert B., Michael Heath, and Christopher P. Lynam. 2023. 'Can We Use Recovery Timescales to Define Good Environmental Status?' *Ecological Indicators* 155 (November): 110984. https://doi.org/10.1016/j.ecolind.2023.110984.

Ulanowicz, Robert E. 1986. *Growth and Development*. Springer. https://doi.org/10.1007/978-1-4612-4916-0.

Vader, Anna, Eleanor Handler, Ragnheid Skogseth, Aud Larsen, and Tove M. Gabrielsen. 2025. 'Seasonality and Interannual Variability of an Arctic Marine Time Series, IsA'. *Arctic Science* 11 (January): 1–17. https://doi.org/10.1139/as-2024-0044.

Varpe, Øystein. 2012. ‘Fitness and Phenology: Annual Routines and Zooplankton Adaptations to Seasonal Cycles’. *Journal of Plankton Research* 34 (4): 267–76. https://doi.org/10.1093/plankt/fbr108.

Walker, Brian, David Salt, and Walter Reid. 2006. ‘Resilience Thinking: Sustaining Ecosystems and People in A Changing World’. *Bibliovault OAI Repository, the University of Chicago Press*, January 1.

Whitehouse, George A., Troy W. Buckley, and Seth L. Danielson. 2017. ‘Diet Compositions and Trophic Guild Structure of the Eastern Chukchi Sea Demersal Fish Community’. *Deep Sea Research Part II: Topical Studies in Oceanography* 135 (January): 95–110. https://doi.org/10.1016/j.dsr2.2016.03.010.

Wickham, Hadley. 2016. *Ggplot2: Elegant Graphics for Data Analysis*. Second edition. With Carson Sievert. Use R! Springer international publishing.

Windnagel, A, M Brandt, F Fetterer, and W Meier. 2017. ‘Sea Ice Index Version 3 Analysis’. *NSIDC Special Report* 19.

Yool, A., E. E. Popova, and T. R. Anderson. 2013. ‘MEDUSA-2.0: An Intermediate Complexity Biogeochemical Model of the Marine Carbon Cycle for Climate Change and Ocean Acidification Studies’. *Geoscientific Model Development* 6 (5): 1767–811. https://doi.org/10.5194/gmd-6-1767-2013.